\documentclass[12pt,preprint]{aastex}
\usepackage{graphicx}
\bibpunct{(}{)}{;}{a}{}{,}

\slugcomment{nothing}	
\shorttitle{The cyclotron resonance of A0535+262 at a low luminosity}

\shortauthors{Terada et al.}

\begin{document}

\title{Cyclotron resonance energies at a low X-ray luminosity:\\
A0535+262 observed with {\it Suzaku}}

\author{ 
Y. Terada\altaffilmark{1},T. Mihara\altaffilmark{1},
M. Nakajima\altaffilmark{1}, 
M. Suzuki\altaffilmark{1}, N. Isobe\altaffilmark{1},
K. Makishima\altaffilmark{1,2}, \\H. Takahashi\altaffilmark{2},
T. Enoto\altaffilmark{2}, M. Kokubun\altaffilmark{2}, 
T. Kitaguchi\altaffilmark{2},
S. Naik\altaffilmark{3}, 
T. Dotani\altaffilmark{3}, F. Nagase\altaffilmark{3},\\
T. Tanaka\altaffilmark{3}, S. Watanabe\altaffilmark{3},
S. Kitamoto\altaffilmark{4}, K. Sudoh\altaffilmark{4}, 
A. Yoshida\altaffilmark{5}, Y. Nakagawa\altaffilmark{5}, \\
S. Sugita\altaffilmark{5}, T. Kohmura\altaffilmark{6}, 
T. Kotani\altaffilmark{7},   D. Yonetoku\altaffilmark{8},
L. Angelini\altaffilmark{9}, J. Cottam\altaffilmark{9}, \\
K. Mukai\altaffilmark{9},  R. Kelley\altaffilmark{9},
Y. Soong\altaffilmark{9},
M. Bautz\altaffilmark{10}, S. Kissel\altaffilmark{10}, 
and J. Doty\altaffilmark{11}\\
}

\email{terada@riken.jp}

\altaffiltext{1}{Cosmic Radiation Laboratory, 
  Institute of Physical and Chemical Research,
    Wako, Saitama, Japan 351-0198}
\altaffiltext{2}{Department of Physics, University of Tokyo, 
    7-3-1 Hongo, Bunkyo-ku, Tokyo, Japan 113-0033}
\altaffiltext{3}{Institute of Space and Astronautical Science, JAXA,
    3-1-1 Yoshinodai, Sagamihara, Kanagawa, Japan 229-8510}
\altaffiltext{4}{Department of Physics, Rikkyo University, 
    3-34-1 Nishi-Ikebukuro, Toshima-ku, Tokyo, Japan 171-8501}
\altaffiltext{5}{Department of Physics \& Math., Aoyama Gakuin University, 
    Sagamihara, Kanagawa, Japan 229-8558}
\altaffiltext{6}{Department of Physics, Kogakuin University, 
    2665-1 Nakano, Hachi-oji, Tokyo, Japan 192-0015}
\altaffiltext{7}{Department of Physics, Tokyo Tech, 
    2-12-1 O-okayama, Meguro-ku, Tokyo, Japan 152-8551}
\altaffiltext{8}{Department of Physics, Kanazawa University,
    Kadoma, Kanazawa, Ishikawa, Japan 920-1192}
\altaffiltext{9}{Exploration of the Universe Division, Code 660, 
NASA/GSFC, Greenbelt, MD 20771, USA}
\altaffiltext{10}{Kavli Institute for Astrophysics and Space Research, 
Massachusetts Institute of Technology, 
77 Massachusetts Avenue, Cambridge, MA 02139}
\altaffiltext{11}{Noqsi Aerospace, Ltd., 2822 South Nova Road,
Pine, Colorado,  80470, USA}


\begin{abstract}

The binary X-ray pulsar A0535+262 was observed
with the {\it Suzaku} X-ray observatory,
on 2005 September 14  for a net exposure of 22 ksec.
The source was in a declining phase of a minor outburst,
exhibiting  3--50 keV luminosity of 
$\sim 3.7 \times 10^{35}$ ergs s$^{-1}$
at an assumed distance of 2 kpc.
In spite of the very low source intensity 
(about 30 mCrab at 20 keV),
its electron cyclotron resonance was detected
clearly  with the {\it Suzaku} Hard X-ray Detector,
in absorption at about 45 keV.
The resonance energy is found to be
essentially the same as those measured
when the source is almost two orders of magnitude more luminous.
These results are compared with the luminosity-dependent
changes in the cyclotron resonance energy,
observed from 4U\,0115+63 and X0331+53.
\end{abstract}

\keywords{magnetic fields --- pulsars: individual (A0535+26) 
--- stars:neutron --- X-rays: binaries}

\section{Introduction}
\label{section:introduction}
Binary pulsars are considered to have strong surface 
magnetic fields in the range of $10^{12-13}$\,Gauss.
Their field strengths can be directly measured through energies 
of cyclotron resonance scattering features (CRSFs),
which often appear in their X-ray spectra (e.g Mihara 1995; Makishima et al. 1999).
Neglecting the gravitational redshift, 
the magnetic field strength $B$ is then calculated 
from the resonance energy $E_{\rm a}$ through a relation of
$ E_{\rm a}\,({\rm keV}) = 11.6\  B/ (10^{12}\,{\rm Gauss}) $. 

Since the magnetic field must be intrinsic to a pulsar, 
the CRSF energy was  believed to be constant in each object.
However, a $\sim$35 \% change in the cyclotron energy 
was unexpectedly observed with {\it Ginga} 
from the recurrent transient pulsar 4U\,0115+63,
between its 1990 and 1991 outbursts  (Mihara 1995; Mihara et al.\ 1998, 2004).
The change is considered to reflect luminosity-dependent 
variations in  the accretion-column height 
by several hundred meters,
assuming a  dipolar field configuration (Mihara et al. 2004).
The effect was studied in further detail by Nakajima et al.\ (2006),
using the {\it RXTE} data of 4U\,0115+63 
which continuously covered another outburst in 1999.
At that time, the CRSF energy was confirmed to increase from $\sim$11 keV 
to $\sim$16 keV,
as the 3--50 keV source luminosity (at an assumed distance of 7 kpc;
Negueruela \& Okazaki 2001) decreased across
a relatively narrow range of $(2-4) \times 10^{37}$ ergs s$^{-1}$.

From another source, X0331+53 (V0332+53), 
a similar effect was detected  with {\it INTEGRAL} (Mowlavi et al.\ 2006) 
and {\it RXTE} (Nakajima 2006b). 
The change in the CRSF energy, however, started in this case 
at a higher luminosity of  $2 \times 10^{38}$ ergs s$^{-1}$,
assuming this object to have a distance of 7 kpc as well 
(Negueruela et al. 1999). 
Thus, the luminosity-dependent change in the CRSF energy 
is emerging as a new intriguing issue,
of which a unified view is yet to be constructed.

This {\it Letter} deals with  A0535+262,
yet another recurrent transient  
with a 103 sec pulsation period and a 111 day orbital period,
located at a distance of 2.0 kpc (Steele et al. 1998).
Its CRSFs were discovered  at 50 and 100 keV 
by the TTM and HEXE instruments
onborad {\it Mir-KVANT} \/ in a 1989 outburst (Kendziorra et al.\ 1994),
and the 2nd harmonic was later reconfirmed
at 110 keV with the {\it CGRO} OSSE (Grove et al. 1995).
Since this pulsar has the {\em highest}  measured CRSF energy,
it is of particular interest to compare this object with 4U~0115+53, 
which has the {\em lowest}  known CRSF energy.
However, the previous CRSF measurements  from A0535+262 
were all limited to very luminous states.
Here, we report on a successful detection of the CRSF from this source 
by {\it Suzaku}, made for the first time 
at a very low luminosity of $\sim 4 \times 10^{35}$ ergs s$^{-1}$ (Inoue et al. 2005).

\section{Observations and Data Reduction}
\label{section:observation}

According to {\it RXTE} ASM monitoring, 
A0535+262 entered outburst twice in 2005.
On the first occasion,
the 2--10 keV intensity reached a peak of 1.3 Crab
($\sim 1 \times 10^{38}$ ergs s$^{-1}$ in luminosity)  on June 6.
The second outburst, an order of magnitude smaller, 
took place about one binary orbital period later, 
reaching the peak  on September 1.
On 2005 August 28
when the  2--100 keV luminosity was $ 1.2 \times 10^{37} $ ergs s$^{-1}$,
the CRSF was detected at $48.5 \pm 0.7$ keV with the {\it RXTE}  (Wilson \& Finger 2005).
The same feature was confirmed with the {\it INTEGRAL} SPI
at $47 \pm 2$ keV on August 31 (Kretschmar et al.\ 2005). 
These measurements refer to Gaussian modeling of the observed 
absorption features.

The fifth Japanese X-ray satellite, {\it Suzaku}, was launched on 2005 July 10.
It carries onboard the X-ray Imaging Spectrometer (XIS) 
operating in 0.2--12 keV (Matsumoto et al. 2005), 
and the Hard X-ray Detector (HXD; Kawaharada et al.\ 2004)
which covers 10--70 keV with PIN diodes and 40--600 keV with GSO scintillators.

We observed A0535+262 with {\it Suzaku} 
from 13:40 UT on 2005 September 14 to 01:00 UT the next day,
when the source was in the declining phase of the second outburst.
The observation was carried out at  ``XIS nominal''  pointing position,
for a net exposure of 22.3 ksec with the XIS and 21.7 ksec with the HXD.
The XIS was operated in the normal mode with ``1/8 window'' option,
which gives a time resolution of 1 sec, 
whereas the HXD was in the nominal mode.

The source was detected 
with the XIS at an intensity of 10 cts s$^{-1}$ per sensor.
In the XIS analysis, 
we excluded all the telemetry saturated data portions,
and  data taken in ``low" rate mode.
We further removed those time intervals 
when  the source elevation above the earth's limb was below $5^{\circ}$
or  the spacecraft was  in the South Atlantic Anomaly (SAA).
We then accumulated nominal-grade events 
within 6 mm ($4'.3$) of the image centroid.
The XIS background spectra were taken 
from a blank sky observation towards the North Ecliptic Pole region,
conducted for  95 ksec on 2005 September 2--4.

The HXD data were screened using the same criteria as for the XIS. 
In addition, we discarded data  taken up to 436 s after leaving the SAA,
and those acquired during time intervals where the geomagnetic 
cutoff rigidity was lower than 8 GeV/c. 
After this filtering, the final HXD event list was obtained  only using events 
that survived the standard anti-concidence function of the HXD.

The non X-ray background of the PIN diodes
was synthesized by appropriately combining night-earth data sets
acquired under different conditions (Kokubun et al. 2006).
The GSO background was derived from a source-free observation
performed on 2005 September 13, namely, 
immediately before the pointing onto A0535+26.
The background GSO events were accumulated 
over identical orbital phases of {\it Suzaku} as the on-source data 
integration, resulting in an exposure of 19 ksec.
Though rather short, this particular dataset is considered best 
in minimizing systematic errors associated with the GSO background estimation.
After subtracting these backgrounds, the source was detected significantly 
at an intensity of 1.4 cts s$^{-1}$ (10--70 keV) with PIN,
and 0.4 cts s$^{-1}$ (40--200 keV) with GSO.

\begin{figure}[hbt]
\epsscale{1.2}
\plotone{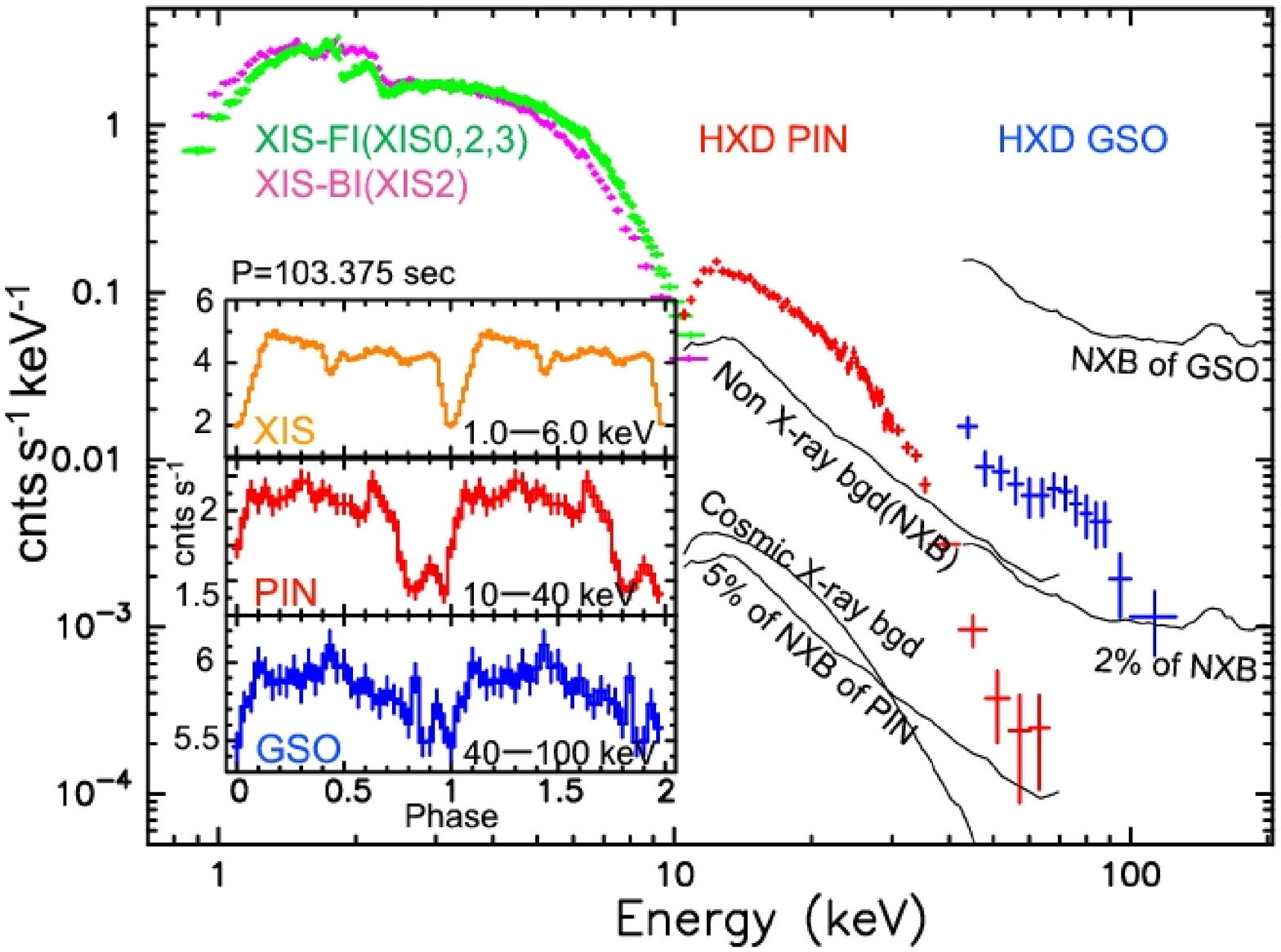}
\caption{Time-averaged and background-subtracted spectra of A0535+262, 
obtained with the {\it Suzaku} XIS (green and purple) and HXD (red and blue).
They are presented without removing instrumental responses, and 
error bars are statistical only. 
Non X-ray backgrounds of PIN and GSO, their uncertainties (see text),
and the estimated cosmic X-ray background (Boldt 1987) are 
shown in solid lines.
The inset shows background-inclusive pulse profiles 
in three typical energy bands, 
obtained by folding the data at the pulsation period.
The relative timings between XIS and HXD are not calibrated yet.
}
\label{fig:a0535_spec_av}
\vspace*{-0.3cm}
\end{figure}

\section{Results}
Figure \ref{fig:a0535_spec_av} shows 
the background-subtracted XIS and HXD spectra of A0535+262,
averaged over the whole observation.
The 0.5 -- 100 keV X-ray flux is measured as
$9.4 \times 10^{-10}$ ergs s$^{-1}$ cm$^{-2}$, 
yielding an X-ray luminosity of 
$4.5 \times 10^{35}$ ergs s$^{-1}$ in the same band.
In Figure~ \ref{fig:a0535_spec_av},
we also show typical PIN and GSO non X-ray backgrounds, 
and uncertainties in their reproducibility
which are $\sim \pm 5\%$ for PIN and $\sim \pm 2\%$ for GSO;
both these estimates are specific to the present observing conditions,
and are based on the {\em current} level of instrumental calibration.
Thus, the statistical errors are dominant in the PIN spectrum,
whereas the statistical and systematic errors are comparable
in the GSO data particularly above 100 keV.
The contribution of the cosmic X-ray background is only $\sim 3\%$ 
of the signal, and hence negligible, but it was
subtracted in deriving the PIN spectrum in Figure \ref{fig:a0535_spec_av}.

The source pulsations were detected 
at a barycentric period of 103.375 $\pm$ 0.09 sec, 
over the full XIS band and at least up to 100 keV with the HXD.
As shown in Figure \ref{fig:a0535_spec_av} inset,
the pulse profiles are similar to those previously
obtained with {\it Ginga} when the source was 80 times brighter (Mihara 1995), 
although the dip around phase 0.45, 
probably caused by absorption, is shallower.

\begin{figure}[htb]
\epsscale{1.1}
\plotone{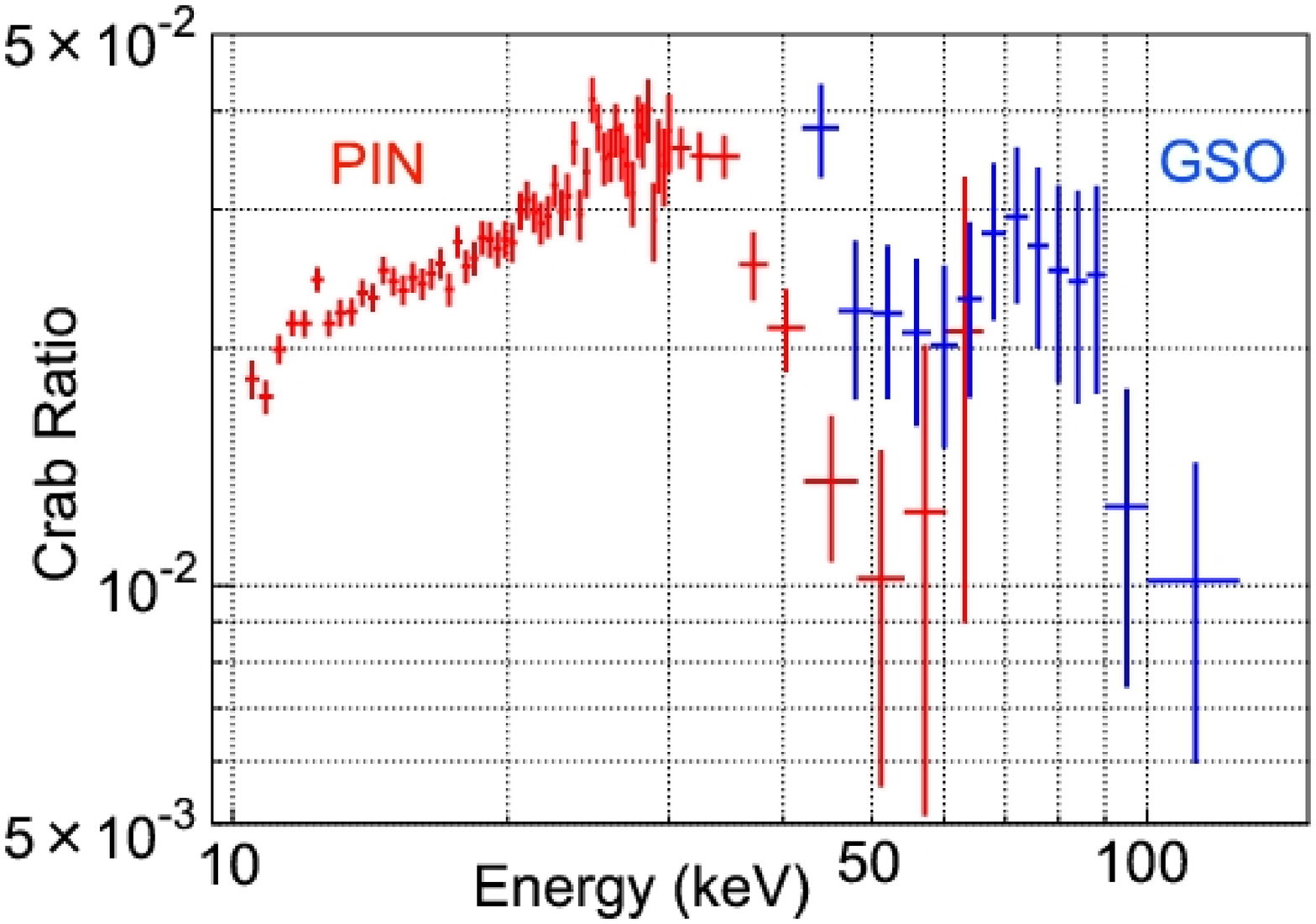}
\caption{
Ratios of the background-subtracted HXD spectra of A0535+262
to those of the Crab Nebula. 
The data obtained with the HXD PIN and GSO 
are shown in red and blue, respectively.
Error bars are statistical only.
}
\label{fig:crabratio}
\end{figure}

In order to evaluate the HXD spectra in a model-independent manner,
we normalized them to those of the Crab Nebula,
acquired in the same detector conditions on 2005 September 15,
immediately after the A0535+262 observation.
The resulting  ``Crab ratios'',
presented in Figure~\ref{fig:crabratio}, indicate that
the source intensity is  $\sim 30$ mCrab at 20 keV.
The ratio keeps rising up to $\sim 30$ keV, and falls steeply beyond,
where the PIN and GSO data both reveal a clear dip feature centered 
at $\sim 50$ keV.
Since the Crab spectrum is a featureless power-law 
with a photon index of $\sim 2.1$,
we may identify this feature with the CRSF of A0535+262 
observed previously (\S~\ref{section:introduction}).
Although the Crab ratios appear somewhat
discrepant between the two HXD components,
the effect can be partially explained by different energy resolutions 
between PIN ($\sim$ 3 keV  FWHM) and GSO ($\sim 10$ keV FWHM at $\sim 50$ keV),
coupled with the steeply declining spectra.
The remainder is within the statistical plus systematic uncertainties.

To quantify the inference from Figure~\ref{fig:crabratio},
we jointly fitted the PIN and GSO spectra,
with the smooth continuum model called NPEX
(Negative and Positive power-laws with EXponential;
Mihara 1995; Makishima et al. 1999).
We left free all but one parameters describing the model;
two normalizations, the negative power-law index ($\alpha_1$),
and the exponential cutoff parameter ($kT$).
The positive power-law index, $\alpha_2$, was fixed at 2.0,
representing a Wien peak, because it was not well constrained by the fit.
The interstellar absorption was not
incorporated, and the relative PIN vs. GSO normalization
was fixed to 1.0 (nominal value).
We limited the GSO fitting range to 50--100 keV,
because of response uncertainties below 50 keV
and the systematic background errors above 100 keV
(Fig. \ref{fig:a0535_spec_av}).

\begin{figure}[hbt]
\epsscale{1.1}
\plotone{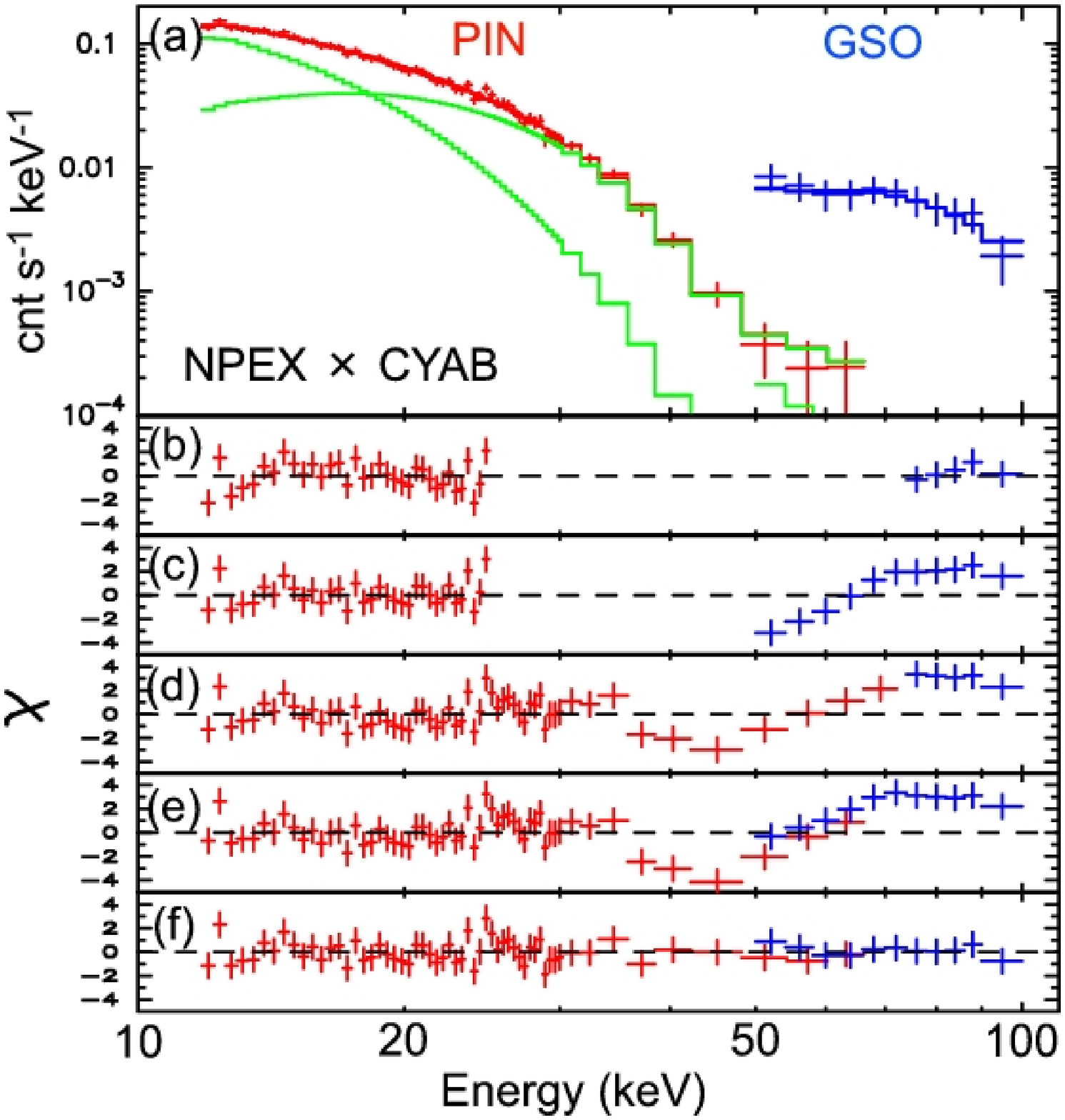}
\caption{
Model fittings to the pulse phase-averaged HXD spectra of A0535+262.
Panel (a) shows the NPEX $\times$ CYAB fit
in comparison with the data, and panel (f) the data-to-model ratio. 
In panels (b) through (e),
the PIN and GSO data are divided by the
best-fit NPEX models (without incorporating the CYAB factor)
obtained using different energy ranges (see text).
}
\label{fig:fit}
\end{figure}

As presented in Figure \ref{fig:fit}b,
the NPEX model was successful with $\chi_\nu^2=1.23$  ($\nu=36$)
when fitted to the 12--25 keV PIN data 
and the 73--100 keV GSO data
(i.e., excluding the suggested CRSF energy range).
When the 50--73 keV GSO data are restored,
a worse fit ($\chi_\nu^2=1.92, \nu=42$) was
obtained as shown in Figure \ref{fig:fit}c.
Similarly, inclusion of the 25--70 keV PIN data
(but discarding the 50--73 keV GSO data) resulted in a poor fit, 
with $\chi_\nu^2=2.13$ for  $\nu=59$ (Figure \ref{fig:fit}d).
The fit became completely unacceptable 
($\chi_\nu^2=2.51$ with  $\nu=65$; Figure \ref{fig:fit}e)
when the entire 12--70 keV PIN data 
and the 50--100 keV GSO spectra are utilized.
Thus, the PIN and GSO data consistently indicate
the spectral feature between $\sim 25$ and $\sim 70$ keV.
This conclusion does not change even considering 
the systematic background uncertainties (Figure~\ref{fig:a0535_spec_av}).

To better describe the HXD spectra,
we multiplied the NPEX continuum 
with a cyclotron absorption (CYAB) factor
(Mihara et al.\ 1990; Makishima et al. 1999),
and repeated the joint fitting.
We left free the energy $E_{\rm a}$, 
depth $D$, and width $W$ of the resonance.
As presented in Figures~\ref{fig:fit}a and f,
this NPEX$\times$CYAB model is successful in reproducing
the PIN (12--70 keV) and the GSO (50--100 keV) spectra simultaneously,
with $\chi^2_\nu = 0.89$ for $\nu=61$.
The CYAB parameters were constrained as  
$E_{a} = 45.5^{+1.3}_{-1.3} $$ ^{+0.9}_{-0.7}$ keV,
$D= 1.8^{+0.4}_{-0.3} $$ ^{+0.5}_{-0.2}$,
and $W = 10.9^{+4.2}_{-2.9} $$^{+2.1}_{-0.7}$ keV,
where the first ``$\pm$" represent statistical 90\% errors
while the second ones show the effect of systematic background uncertainties.
Compared with these, systematic errors 
in the energy scale determination have smaller effects;
the PIN energy scale, reconfirmed by Gd-K lines at 43 keV, is accurate to within 1 keV, 
and that of GSO, calibrated by several instrumental lines, 
is reliable to within a few keV.
In addition, the systematics in the response matrix are 
also within statistical errors in the present analysis.
The NPEX parameters became
$\alpha_1 = -1.62^{+0.44}_{-0.31} $$^{+0.04}_{-0.02}$,
and $kT= 12.2^{+0.8}_{-0.7}$$^{+0.7}_{-1.3}$ keV, with 
the positive and negative power-law components crossing over 
at $\sim 20$ keV (green curves in Fig.~\ref{fig:fit}).
This successful NPEX$\times$ CYAB fit reinforces
our identification of the feature with the previously known CRSF.

If we instead adopt a ``power-law times exponential'' continuum 
with  Gaussian absorption  (GABS; Kreykenbohm et al.\ 2004),
the HXD data give $E_{\rm a} = 46.3^{+1.5}_{-1.3} $$^{+1.9}_{-1.0}$ keV,
a Gaussian sigma of $= 4.4^{+1.0}_{-0.9} $$^{+0.6}_{-0.3}$ keV,
and an optical depth of $2.2^{+0.9}_{-0.7} $$^{+1.8}_{-0.3}$.
However, the fit becomes worse; $\chi^2_\nu = 1.54$ for $\nu=63$.

Our data are consistent with the presence of the 2nd harmonic feature 
at at $\sim 100$ keV (see Fig \ref{fig:a0535_spec_av}),
but do not require it,
given the current status of the HXD calibration.
If the model is multiplied by another CYAB factor 
with the center energy and width fixed at $2E_{\rm a}$ and $2W$, respectively,
its depth is constrained as $0 \leq D_2 < 1.9$.

\section{Discussion}

We observed A0535+262 with {\it Suzaku} 
in a very low luminosity state,
$4.5 \times 10^{35}$ ergs s$^{-1}$ in 0.5--100 keV,
or $3.7 \times 10^{35}$ ergs s$^{-1}$ in 3--50 keV.
In spite of the source faintness ($\sim$30 mCrab),
we successfully detected  the CRSF at $\sim 45$  keV,
thanks to the wide energy band and high sensitivity of the HXD.
The CRSF was as deep as (or even deeper than) 
that in the high luminosity states;
e.g., $D \sim 0.5$ when the source was two orders of magnitude
more luminous (Kendziorra et al. 1994; Grove et al. 1995).

Except for  the case of X Persei (Coburn et al.\ 2001),
which exhibits a rather unusual spectrum for an accreting pulsar,
the present result  provides the detection of a CRSF
in the lowest luminosity state ever achieved from a binary X-ray pulsar.
Since the CRSF appeared in absorption and not in emission,
the Thomson optical depth of the accretion column 
is inferred to be larger than $\sim$10
even with this low luminosity,
according to the Nagel model (1981).
This conclusion is independently supported by the 
fact that the pulse profiles (Figure~\ref{fig:a0535_spec_av})
do not differ significantly
from those observed in much more luminous outbursts.

\begin{figure}[htb]
\bigskip
\epsscale{1.0}
\plotone{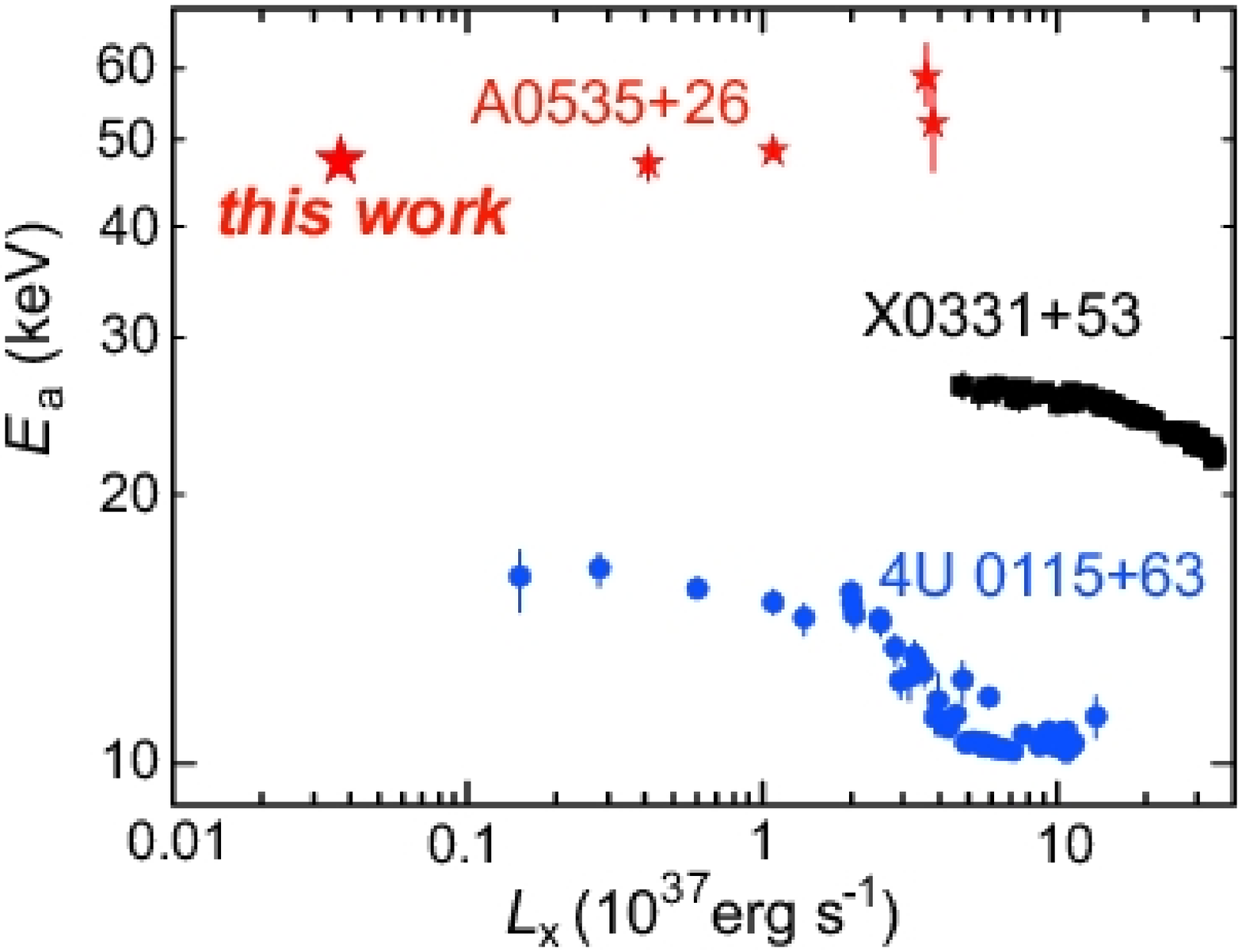}
\caption{
Luminosity (3--50 keV) dependence of the fundamental cyclotron resonance 
energies in binary X-ray pulsars,
using Gaussian absorption modeling. 
The other data points for A0535+262 refer to 
Wilson \& Finger (2005; $L_{\rm x}=1.1 \times 10^{37}$ ergs s$^{-1}$), 
Kretschmar et al.\ (2005; $0.4 \times 10^{37}$), 
Grove et al.\ (1995; $ 3.6 \times 10^{37}$, 
assuming 110 keV as the 2nd harmonic),
and  Kendziorra et al. (1994; $ 3.8 \times 10^{37}$).
The results on 4U\,0115+63 and X0331+53,
both assuming a distance of 7 kpc,
are from Nakajima et al.\ (2006) and  Nakajima (2006b), respectively.}
\label{fig:nakajimaplot}
\end{figure}

In Figure \ref{fig:nakajimaplot},
we plot the  CRSF energy of A0535+262
versus the 3--50 keV  luminosity at 2 kpc, $L_{\rm x}$, 
as measured by {\it Suzaku} and by previous missions.
For consistency with previous works, 
the {\it Suzaku} result plotted here is from the GABS modeling.
The cyclotron resonance energy of A0535+262 
is  thus constant within $\pm 10\%$,
even though the luminosity changed by nearly  two orders of magnitude
from $4 \times 10^{35}$ to $4 \times 10^{37}$  ergs s$^{-1}$.
Therefore,  the implied magnetic field strength, 
$4.0 \times 10^{12}$ Gauss,
is considered to represent a value intrinsic to this object, 
which is most likely  that on the neutron star surface.

Figure \ref{fig:nakajimaplot} also presents results 
on the other two pulsars, 4U\,0115+63 and X0331+53, 
mentioned in \S~\ref{section:introduction}.
The three objects behave in rather different ways
on the $L_{\rm x}$ vs. $E_{\rm a}$ plane.
While 4U\,0115+63 and X0331+53 both exhibit
luminosity-dependent changes in $E_{\rm a}$,
the threshold luminosity at which the resonance energy 
starts changing is significantly different.
In addition, A0535+262 does not show such behavior
at least up to $\sim 4 \times 10^{37}$ erg s$^{-1}$.

Considering that accreting X-ray pulsars should form a  family
described by a  rather small number of parameters,
one possibility suggested by Figure \ref{fig:nakajimaplot} is
that the $L_{\rm x}$ vs. $E_{\rm a}$ relation depends 
systematically on some of the parameters,
e.g., the surface field strength itself.
Comparing 4U\,0115+63 and X0331+53,
we may speculate that $E_{\rm a}$  starts changing at a higher luminosity
if the object has a higher surface magnetic field.
If this is correct, 
we would expect the resonance energy of A0535+262 
to change at luminosities much higher than so far sampled, 
because  it has the highest magnetic field among known CRSF pulsars.

Alternatively,  the values of $L_{\rm x}$ in 
Figure~ \ref{fig:nakajimaplot} may be  subject to systematic errors,
due, e.g.,  to uncertainties in the source distance,
and/or  to the anisotropy of emission that is inherent to X-ray pulsars.
(The luminosities are all calculated assuming isotropic emission.)
If the luminosity is corrected for these factors,
the behavior of $E_{\rm a}$ in Figure~\ref{fig:nakajimaplot}
might become essentially the same among the three objects.

To distinguish between these two possibilities
(or to arrive at yet another alternative),
we need further observations.
In either case,  it must be examined 
whether the variable column-height scenario,
which was successful on 4U\,0115+63 
(Mihara et al. 1998, 2004; Nakajima et al. 2006),
can be applied also to  A0535+262 and X0331+53.

\acknowledgements
The authors would like to thank  all the members of 
the {\it Suzaku} Science Working Group,
for their contributions in the instrument preparation,
spacecraft operation, software development,
and in-orbit instrumental calibration.



\end{document}